\documentclass[12pt]{iopart}
\usepackage{natbib}

\usepackage[dvips]{graphicx}
\usepackage[dvips,usenames]{color}

\newcommand{\avk}{\left< k \right>}

\newcommand{\al}{\alpha}

\renewcommand{\pt}{\partial_t}

\begin{document}

\letter{Effects of mobility on ordering dynamics }

\author{Andrea Baronchelli and  Romualdo Pastor-Satorras}

\address{Departament de F\'\i sica i Enginyeria Nuclear, Universitat
  Polit\`ecnica de Catalunya, Campus Nord B4, 08034 Barcelona, Spain}

\begin{abstract}
    Models of ordering dynamics allow to understand
    natural systems in which an initially disordered population
    homogenizes some traits via local interactions.  The simplest of
    these models, with wide applications ranging from evolutionary to
    social dynamics, are the Voter and Moran processes, usually
    defined in terms of static or randomly mixed individuals that
    interact with a neighbor to copy or modify a discrete trait.  Here
    we study the effects of diffusion in Voter/Moran processes by
    proposing a generalization of ordering dynamics in a
    metapopulation framework, in which individuals are endowed with
    mobility and diffuse through a spatial structure represented as a
    graph of patches upon which interactions take place. 
    We show that diffusion dramatically affects the time to
    reach the homogeneous state, independently of the underlying
    network's topology, while the final consensus emerges through
    different local/global mechanisms, depending on the mobility
    strength. Our results highlight the crucial role played by mobility
    in ordering processes and set up a general framework that allows to
    study its effect on a large class of models, with implications in
    the understanding of evolutionary and social phenomena.
\end{abstract}

\pacs{87.23.-n, 05.40.-a,89.75.-k}

\maketitle

Initially heterogeneous individuals in isolated groups often end up
homogenizing their characteristic traits: People tend to align their
opinions \cite{castellano2007sps}, segregated populations gradually
lose their genetic diversity \cite{westemeier1998tlt},  different social
groups spontaneously develop their own seemingly arbitrary traits,
such as distinctive styles of dress or jargons \cite{barth1969ega}, etc.  In all
these cases a global order emerges trough local interactions in a
self-organized way, without any central coordination.  Several
statistical models have been put forward to capture the main features
of this kind of ordering processes.  Among the simplest ones range the
Voter model (VM) \cite{liggett1985ips} and the Moran process (MP)
\cite{moran1962spe}, designed to address issues of social
\cite{castellano2007sps} and evolutionary \cite{nowak2006ed} dynamics,
respectively. In the VM, individuals hold one of two mutually
exclusive opinions, and are subject to pairwise interactions in which
a randomly selected individual adopts the opinion of a nearest
neighbor peer. In the MP, on the other hand, individuals
belong to one (out of two) species and reproduce generating an equal
offspring which replaces a randomly selected nearest neighbor.  Due to
their abstraction and simplicity, both models are now well established
paradigms of ordering dynamics \cite{castellano2007sps,nowak2006ed}.

Previous statistical studies of VM/MP dynamics have mainly considered
the randomly mixed \cite{moran1962spe} (mean-field) case or a static
(\textit{fermionic}) distribution of individuals, identified with the
sites of a lattice \cite{castellano2007sps}. More recently, after the
discovery that the topological environment of many social and
ecological processes is highly heterogeneous \cite{barabasi02},
fermionic dynamics on complex networks have also been considered
\cite{Castellano03,lieberman2005edg,sood05:_voter_model}.  While
interesting results have been obtained in all these cases, the
analysis of the effects of the \textit{mobility} of individuals has
been mostly neglected, even thought it is a crucial feature of many
real biological and social systems. For example, human migration
guarantees cultural contamination \cite{sjaastad1962car}, while small
exchanges between separated groups yield  spatially synchronized
population oscillations \cite{blasius1999cda}, and mobility of
individuals promotes biodiversity \cite{reichenbach2007mpa}. In an
evolutionary context, moreover, migration is the force that increases
the inter-population similarity reducing the intra-population
homogeneity, thus contrasting the effects of random drift and
adaptation \cite{rice04:_evolut_theor}.

Here we explore the role of mobility in ordering dynamics by
considering the VM/MP within a generalized metapopulation
(\textit{bosonic}) framework
\cite{v.07:_react,baronchelli08:_boson_react}. As in classical studies
in population genetics \cite{neal04:_introd}, and recent
generalizations \cite{baxter:258701}, we consider individuals of
different species placed on a geographical substrate, represented for
generality in terms of a random graph or network \cite{barabasi02},
whose vertices can host a population of any desired size. Individuals
are endowed with mobility and at each time step they can either
interact with the local population or migrate to a nearest neighbor
vertex. To take mobility quantitatively into account, we introduce a
species-specific parameter, representing the ratio between the
mobility and interaction strengths, that determines the probability
that an individual performs one of these two steps.  We present
evidence that mobility can strongly affect the ordering process,
determining the onset of different mechanisms leading the system to
the final homogeneous state and dramatically affecting the average
time needed to reach it.  Our results imply that the coupling between
mobility and interactions leads to novel properties of the dynamics of
ordering, and should be explicitly taken into account when aiming at
realistic modeling in general social or biological contexts.

In our metapopulation scheme, individuals interact inside the vertices
of a network, while they can diffuse along the edges connecting pairs
of vertices.  From a statistical point of view, the underlying network
is described by means of its degree distribution $P(k)$ (probability
that a vertex has degree $k$---i.e. is connected to $k$ other
vertices) \cite{barabasi02} and its degree-degree correlations
$P(k'|k)$ (conditional probability that a vertex of degree $k$ is
connected to a vertex of degree $k'$) \cite{alexei}.  Individuals
belong in general to $S$ different species, 
each defined by a given trait (opinion, genotype, etc.) $\alpha$ and
characterized by a parameter $p_\alpha$ (the \textit{mobility ratio}),
representing the ratio between the mobility (diffusion coefficient)
and the propensity of species $\al$ to interact with other species.
The dynamics of the processes is defined in the spirit of discrete
time stochastic particle systems \cite{marro1999npt}: At time $t$, one
individual is randomly selected, belonging to class $\alpha$.  With
probability $p_\alpha$, the individual migrates, performing a random
jump to a nearest neighbor vertex.  Otherwise, with probability
$1-p_\alpha$, the individual chooses a peer inside its same vertex
(the peer belonging to the species $\alpha'$) and reacts with it
according to the dynamical rules describing the corresponding model:
\textbf{(i)} Metapopulation VM (MVM): The individual copies the trait
of the peer and becomes of species $\alpha'$.  \textbf{(ii)}
Metapopulation MP (MMP): The individual reproduces, generating an
offspring of the same species $\alpha$, which replaces the peer.  In
any case, time is updated as $t \to t+1/N$, where $N$ is the (fixed)
number of individuals.  For each species $\al$ the occupation number
of any vertex is unbounded and can assume any integer value, including
zero \footnote{We note that, with our definition, the occupation number
  of each vertex is not fixed, as in previous metapopulation models
  \cite{baxter:258701}, but it is in fact a stochastic variable whose
  average value depends in general on the network structure.}. When
$p_{\alpha}=p_{\alpha'} \;\; \forall \; \alpha,\alpha'$ the MVM and
the MMP are obviously equivalent \cite{sood08:_voter_model}.

\begin{figure}[t]
 \begin{center}
   \includegraphics*[width=12cm]{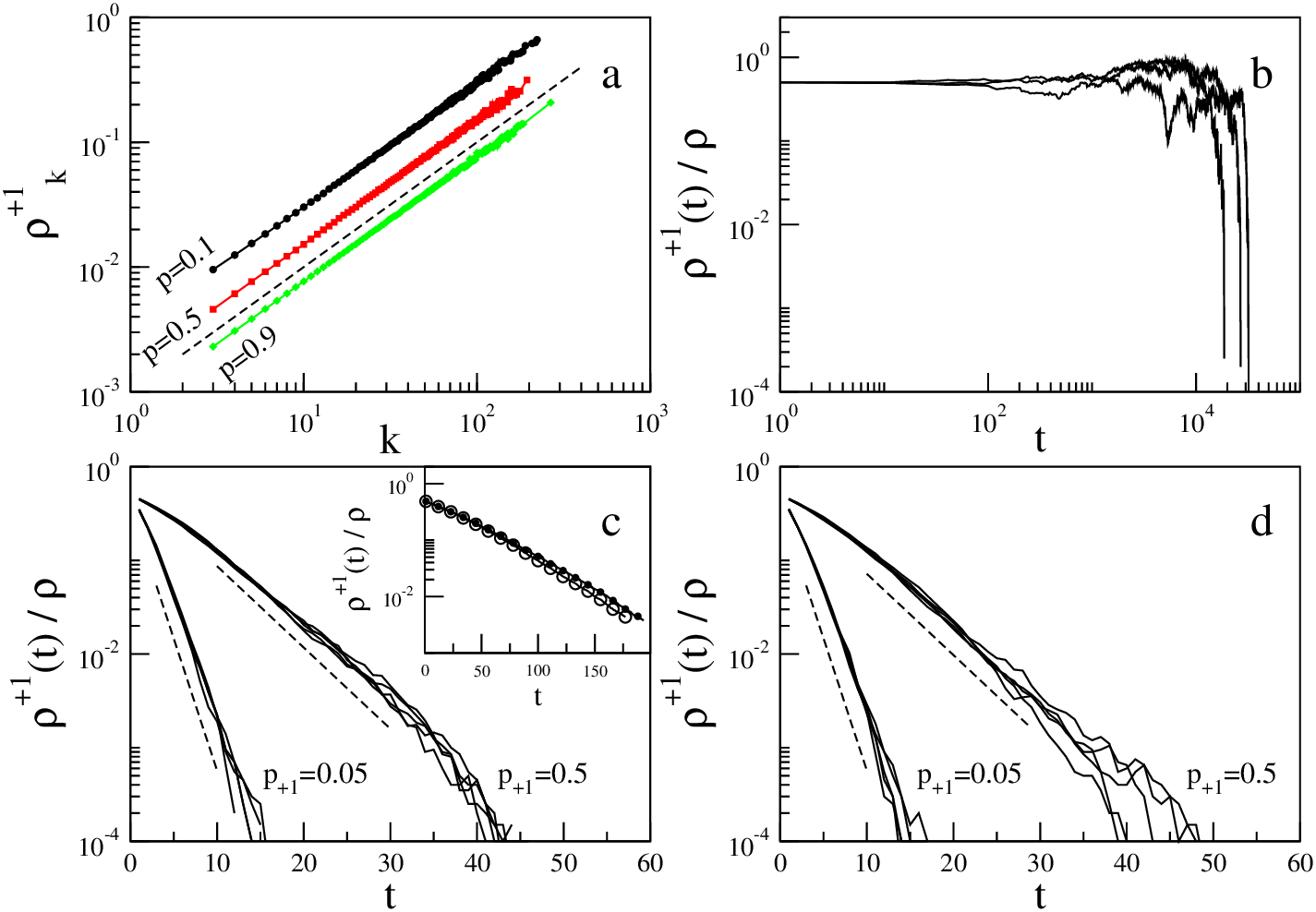}
 \end{center}
 \caption{(a) Density of $+1$ individuals as a function of the degree in
   MVM and MMP on heterogeneous scale-free networks generated with the
   (uncorrelated) configuration model \cite{ucmmodel} with degree
   distribution $P(k) \sim k^{-2.5}$ for different values of
   $p=p_{+1}=p_{-1}$ (curves ares shifted vertically for clarity). 
   (b) Partial density of $+1$ individuals as a function of
   time ($p_{+1}=p_{-1}=0.5$) in scale-free networks.  (c, main) and
   (d) Partial density of $+1$ individuals as a function of time for the MVM
   in scale-free and fully connected networks, respectively. Dashed
   lines represent the theoretical slope $|p_{+1}-p_{-1}|^{-1}$
   ($p_{-1}=0.7$). Data from single runs with homogeneous initial
   conditions $\rho^{+1}(0) = \rho^{-1}(0)=10$, in networks of size
   $V=10^3$. (c, inset) Also in the low mobility case ($p_{+1}=0.05$,
   $p_{-1}=0.1$) averaged curves for fully connected (empty circles)
   and scale free networks (full circles) collapse well to an
   exponential decay.}
 \label{f:density}
\end{figure}

Both dynamics are characterized by the presence of $S$ ordered states
in which all individuals belong to the same species, and interest lies
in studying how the final ordered state is reached
\cite{castellano2007sps,nowak2006ed}.  The relevant quantities are
thus the \textit{fixation probability} (or exit probability)
$\phi_\al$ and the \textit{consensus} or \textit{fixation time}
$\bar{t}_\al$, defined as the probability that a population ends up
formed by all $\al$ individuals and the average time until the
eventual fixation \cite{castellano2007sps}. To gain insight on these
quantities, it is useful to first consider the time evolution of the
density of individuals.  To do so, we consider as usual the partial
densities of individuals of species $\alpha$ in vertices of degree
$k$, defined as \cite{v.07:_react} $\rho^\al_k(t) = n^\al_k(t) / [V
P(k)]$, where $n^\al_k(t)$ is the number of individuals of species
$\alpha$ in vertices of degree $k$, at time $t$, and $V$ is the
network size.  The total density of species $\al$ is then $\rho^\al(t)
= \sum_k P(k) \rho^\al_k(t)$, satisfying the normalization condition
$\sum_\al \rho^\al(t) = \frac{N}{V} \equiv \rho$, being $\rho$ the
total density of individuals in the network. Let us focus on the
simplest case in which only two species are present in the system,
$+1$ and $-1$, with mobility ratios $p_{+1}$ and $p_{-1}$,
respectively. Within a mean-field approximation
\cite{baronchelli08:_boson_react}, we can see that the quantities
$\rho^\al_k(t)$ fulfill the rate equations
\begin{equation}
   \partial_t \rho^\al_k(t) = - p_\alpha  
  \rho^\al_k(t) +  p_\alpha  k \sum_{k'} \frac{P(k|k')}{k'}
  \rho^\al_{k'}(t) + \varepsilon  (p_{-\al} - p_\al)
  \frac{\rho^\al_{k}(t)\rho^{-\al}_{k}(t)}{\rho^{\al}_k(t)+\rho^{-\al}_k(t)},
  \label{eq:3}
\end{equation}
where $\varepsilon$ takes the values $-1$ and $+1$ for the MVM and the
MMP, respectively. In Eq.~(\ref{eq:3}), the first two terms stand for
the diffusion of individuals, while the third one accounts for the
interactions inside each vertex. Since the total number of individuals
is conserved, the latter disappear in the equation for the density of
individuals, $\rho_{k}(t) = \sum_\al \rho^{\al}_k(t)$, that takes the
form of a weighted diffusion equation for the different species
\cite{v.07:_react,baronchelli08:_boson_react}.  A quasi-stationary
approximation, assuming that the diffusion process is so fast that it
can stabilize the particle distribution in a few time steps, leads to
the functional form for the partial densities, $\rho^{\al}_k(t)= k
\rho^{\al}(t)/\avk$. The existence of this diffusion-limited regime
\cite{baronchelli08:_boson_react}, whose presence is confirmed in
numerical simulations (see Fig.~\ref{f:density}(a)), is expected to
hold for not too small values of $p_\al$. This approximation allows to
write the equation for the total species density
\begin{equation}
   \partial_t \rho^\al(t)= 
   \varepsilon  (p_{-\al} -p_\al)  \rho^{\al}(t) 
  [1-\rho^{\al}(t)/\rho], 
    \label{eq:7}
\end{equation}
where we have used 
the normalization condition $\rho^{+1}(t) + \rho^{-1}(t) =\rho$.
Remarkably, this expression is valid for any degree distribution and
correlation pattern.


When $p_\al = p_{-\al}$ the total species density is conserved ($\pt
\rho^\al(t) = 0$), and the ordering process proceeds via density
fluctuations \cite{castellano2007sps}, see
Fig.~\ref{f:density}(b). For $p_\al \neq p_{-\al}$, if $\varepsilon
(p_{-\alpha} -p_{\alpha}) < 0$ the ordered state corresponds to a
population of the $-\alpha$ species.  The time evolution of the
species $\al$ going extinct is given by an exponential decay,
$\rho^{\al}(t) \sim \rho^{\al}(0) \exp(-t|p_{-\al} -p_\al|)$, see
Fig.~\ref{f:density}(c) and (d). Extinction becomes almost sure when
$\rho^{\al}$ takes its minimum value, namely $N^{-1}$. Therefore,
final ordering takes place in a time of the order $\ln N / |p_{-\al}
-p_\al|$, independently of the network structure, for both MVM and
MMP.

In order to obtain information on the fixation probability, we can
take advantage of the topology independence of the MVM and MMP,
evidenced in Eq.~(\ref{eq:7}), and focus on a fully connected network,
in which the particle distribution is homogeneous in all vertices.
Both MVM and MMP can therefore be mapped to a biased one-dimensional
random walk, in which the transition probabilities $p_{n', n}$ from
$n$ individuals of species $+1$ to $n'$ individuals take the form
\begin{equation}
  p_{n+1,n} = A_+\frac{n(N-n)}{N^2}, \quad   p_{n-1,n} =
  A_-\frac{n(N-n)}{N^2},
  \label{eq:1}
\end{equation}
all the rest being zero except $p_{n,n}= 1 - p_{n+1,n}-p_{n-1,n}$, and
where $A_\pm = 1 - [p_{+1} +p_{-1} \pm \varepsilon (p_{+1}
-p_{-1})]/2$. Applying standard stochastic techniques
\cite{ewens04:_mathem_popul,baronchelli:_in_prepar} one recovers the fixation
probability 
\begin{equation}
  \phi_{+1}(\rho^{+1}) = \frac{1-r^{-V \rho^{+1}}}{1-r^{-V \rho}}
  \label{eq:2}
\end{equation}
where $r = A_+ / A_-$.  This result yields a neat evolutionary
interpretation for the MMP. The fixation probability takes indeed the
same form as in the fermionic MP in any undirected underlying network
\cite{lieberman2005edg}, provided the factor $r =
(1-p_{+1})/(1-p_{-1})$ is interpreted as the relative selective
\textit{fitness} \cite{nowak2006ed} of species $+1$ over species $-1$,
defined as the relative number of offspring contributed to the next
generation by both species \footnote{An analogous interpretation can be
  made for a VM dynamics with fitness \cite{sood08:_voter_model}.}.
In the case $p_{+1}=p_{-1}$, corresponding to the limit $r \to 1$, in
which both species are equivalent, we recover $\phi_{+1} = \rho^{+1}$
\cite{nowak2006ed}.  For $p_{+1}\neq p_{-1}$, on the other hand,
homogeneous initial conditions ($\rho^{+1}=1/2$) in the limit of large
$V$ yield $\phi_{+1} \to \Theta[\varepsilon(p_{-1} - p_{+1})]$ where
$\Theta[x]$ is the Heaviside theta function; that is, the population
becomes, as expected from the analysis of the density evolution,
fixated to the species with the largest (smallest) $p_\al$ value for
the MVM (MMP).

An analysis on general networks \cite{baronchelli:_in_prepar} confirms
the results from fully connected ones, and implies that the MMP
represents therefore a rigorous generalization of the classical
evolutionary MP. Moreover, since the same fitness $r$ can be achieved
for different values of the mobility ratios, the metapopulation
framework allows to explore the independent effects of mobility for a
fixed selective advantage. This is particularly explicit in the form
of the fixation time. To compute it, we extend the backwards
Fokker-Plank approach presented in
Refs.\cite{sood05:_voter_model,antal2006edd,sood08:_voter_model} to
the bosonic case. Focusing for simplicity in the case of $p_{+1} =
p_{-1}\equiv p$ \footnote{The general case $p_{+1} \neq p_{-1}$ will
  be considered elsewhere \cite{baronchelli:_in_prepar}.}, we recast
the stochastic processes defined by the MVM and MMP in terms of a
master equation.  The state of the system can be described by the
occupation number vectors $\vec{n}^\al=\{n_{q}^\al\}$,
$q=1,\ldots,k_c$, where $k_c$ is the largest degree in the network,
that allow to keep track of the actual occupation number of the
vertices of different degree.  Transitions from one state to another
can proceed therefore both when an individual diffuses and when it
changes its state.  Thus, defining the vector
$\vec{\delta}_k=\{\delta_{q,k}\}$, the transitions rates due to the
diffusion of an individual from vertex $k'$ to $k$ take the form
\begin{eqnarray*}
  T(\vec{n}^\al+\vec{\delta}_k -\vec{\delta}_{k'}, \vec{n}^{-\al} |
  \vec{n}^{\al}, \vec{n}^{-\al}) = N \frac{p_\al}{\rho}
  P(k') \rho^\al_{k'}P(k|k'),
\end{eqnarray*}
while the transitions rates due to reaction are
\begin{eqnarray*}
  T(\vec{n}^\al\pm\vec{\delta}_k,  \vec{n}^{-\al} \mp \vec{\delta}_{k} |
  \vec{n}^{\al}, \vec{n}^{-\al}) = N \frac{1-p}{\rho} P(k)
  \frac{ \rho_k^\al \rho_k^{-\al}}{ \rho_k^\al + \rho_k^{-\al}}.
\end{eqnarray*}
From these transition probabilities it is straightforward to write the
corresponding master equation, which can then be translated into a
backwards Fokker-Planck equation, by expanding it up to second order
in the inverse network size $V^{-1}$.  Resorting again to the
quasi-stationary condition $\rho_k^\al = k \rho^\al/\avk$, the
backwards Fokker-Planck becomes a function of the densities $\rho^\al$
only, and its different terms can be conveniently simplified.  From
the backwards Fokker-Planck equation, finally, we obtain the 
consensus time, which, as a function of the reduced initial density
$x=\rho^{+1}/\rho$, satisfies the equation \cite{sood05:_voter_model}
\begin{equation}
  4\frac{1-p}{N} x (1-x)
  \frac{\partial^2\bar{t}_{+1}(x)}{\partial x^2 }= -1,
    \label{eq:6} 
\end{equation}
subject to the boundary conditions $ \bar{t}_{+1}(0)=
\bar{t}_{+1}(1)=0$  \cite{baronchelli:_in_prepar}.  Strikingly, this
equation is the same for both MVM and MMP, and again independent of
the topological details of the network, that therefore turn out to be
an irrelevant parameter as far as the asymptotic results for fixation
probability are concerned (provided that the diffusion rates are not
too small, and the quasi-stationary approximation assumed above is
valid \cite{baronchelli:_in_prepar}). The solution of
Eq.~(\ref{eq:6}) is
\begin{equation}
  \bar{t}_{+1}(x) \sim -\frac{N}{1-p}\left[ x \ln
    x + (1-x)\ln (1-x)\right].
\end{equation}
Therefore, for homogeneous initial conditions, $x=1/2$, we obtain a
fixation time scaling as $\bar{t}_{+1}(1/2) \sim N/(1-p)$.
\begin{figure}[t]
 \begin{center}
   \includegraphics*[width=12cm]{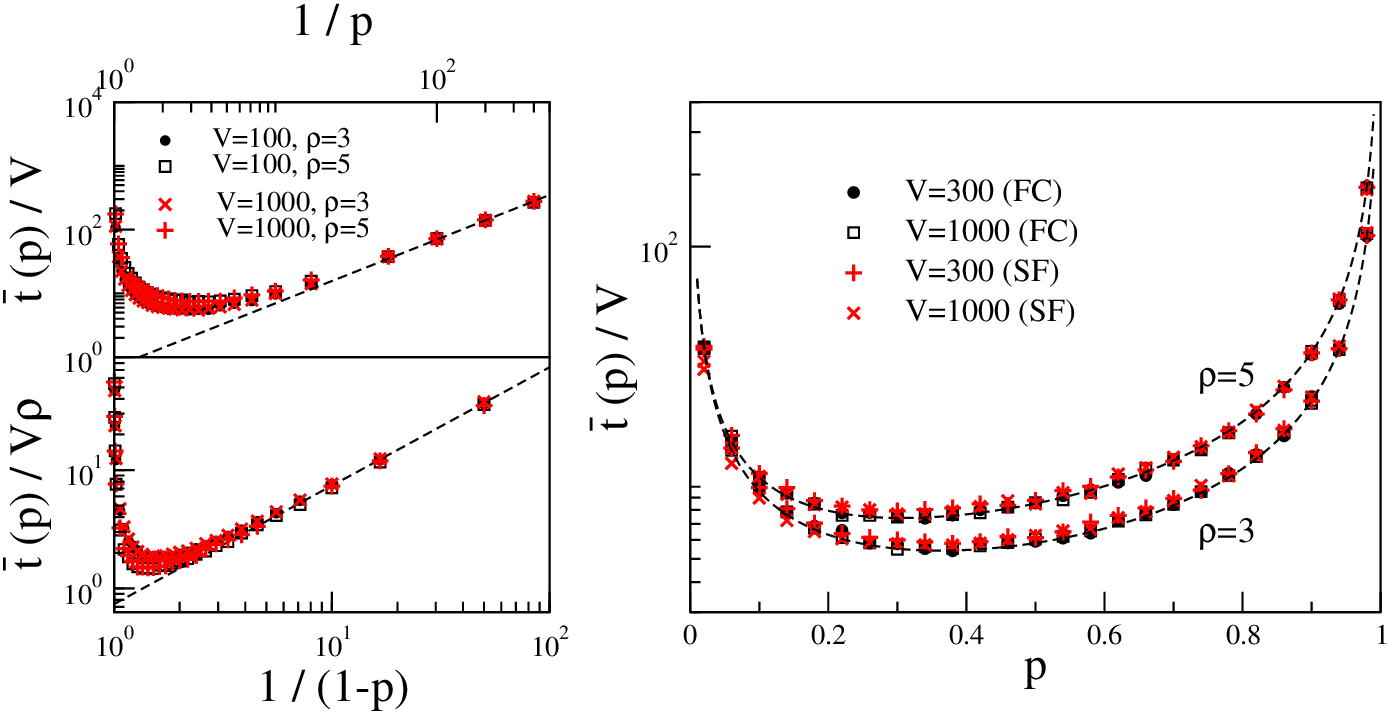}
 \end{center}
 \caption{Left: Scaling of the fixation time for the MVM and MMP with
   mobility $p_{+1} = p_{-1} = p$, in the limit $p \to 0$ (top) and
   $p \to 1$ (bottom), in fully connected networks. Right: Rescaled 
    fixation time for the MVM and MMP with
   mobility $p_{+1} = p_{-1} = p$ in fully connected (FC)
   and scale free (SF) networks of different sizes. Dashed
   lines are nonlinear fits to the functional form Eq.~(\ref{eq:14}),
   for $ A \simeq 0.70$ and $B \simeq 0.72$.  Data refer to
   homogeneous initial conditions.}
 \label{f:fix_t}
\end{figure}
This result recovers the standard scaling linear in $N$ of the
fermionic VM and MP in fully connected networks
\cite{castellano2007sps,nowak2006ed}, and is in opposition to the
topological dependent scaling shown by the VM in heterogeneous
networks \cite{sood08:_voter_model}. The most interesting feature of
this fixation time, however, is that, even though it has been computed
for fixed $r=1$, it shows a strong dependence on the individuals'
mobility $p$. In particular, it is a growing function of $p$, which,
in the limit $p \to 1$ tends to infinity, evidencing a dramatic
slowing down in the ordering process, see Fig.~\ref{f:fix_t}.
Numerical simulations of the fixation time in the full range of
mobility values, Fig.~\ref{f:fix_t} (right panel), yield, however, an
asymmetric concave form, in contrast with the hyperbolic form
predicted by the diffusion approximation. A detailed numerical
analysis, see Fig.~\ref{f:fix_t} (left panel), allows us to conjecture
the functional form of the fixation time as a function of the mobility
$p$, $\bar{t}_{+1}=\bar{t}_{-1}\equiv\bar{t}(p)$, as given by
\begin{equation}
  \bar{t}(p) \approx A\frac{V}{p}+B\frac{V \rho}{1-p},
 \label{eq:14}
\end{equation}
where $A$ and $B$ are constants, approximately independent the
population size and mobility ratio. 
The functional form in Eq.~(\ref{eq:14}) is corroborated by the scaling
analysis performed in Fig.~\ref{f:fix_t} (right panel), were we observe
that curves for fully connected and heterogeneous networks collapse,
when properly scaled, for the same value of $\rho$.  The concave form
of the fixation time implies additionally the presence a minimum for a
value $p_{min}$ of the mobility ratio for which the systems orders
more quickly. According with the estimated functional form in
Eq.~(\ref{eq:14}), this minimum takes the form $p_{min}\sim
\rho^{-1/2}$.
This indicates the striking presence of an optimum global level of
mobility that maximizes the speed at which an opinion consensus is
reached or a neutral mutant dominates a population \cite{nowak2006ed}.
Possibly against intuition, moreover, according to Eq.~(\ref{eq:14}) 
in the thermodynamic limit the fastest
fixation regime is associated to almost still particles ($p_{min} \to
0$ as $\rho \to \infty$).
\begin{figure}[t]
  \begin{center}
    \includegraphics*[width=10cm]{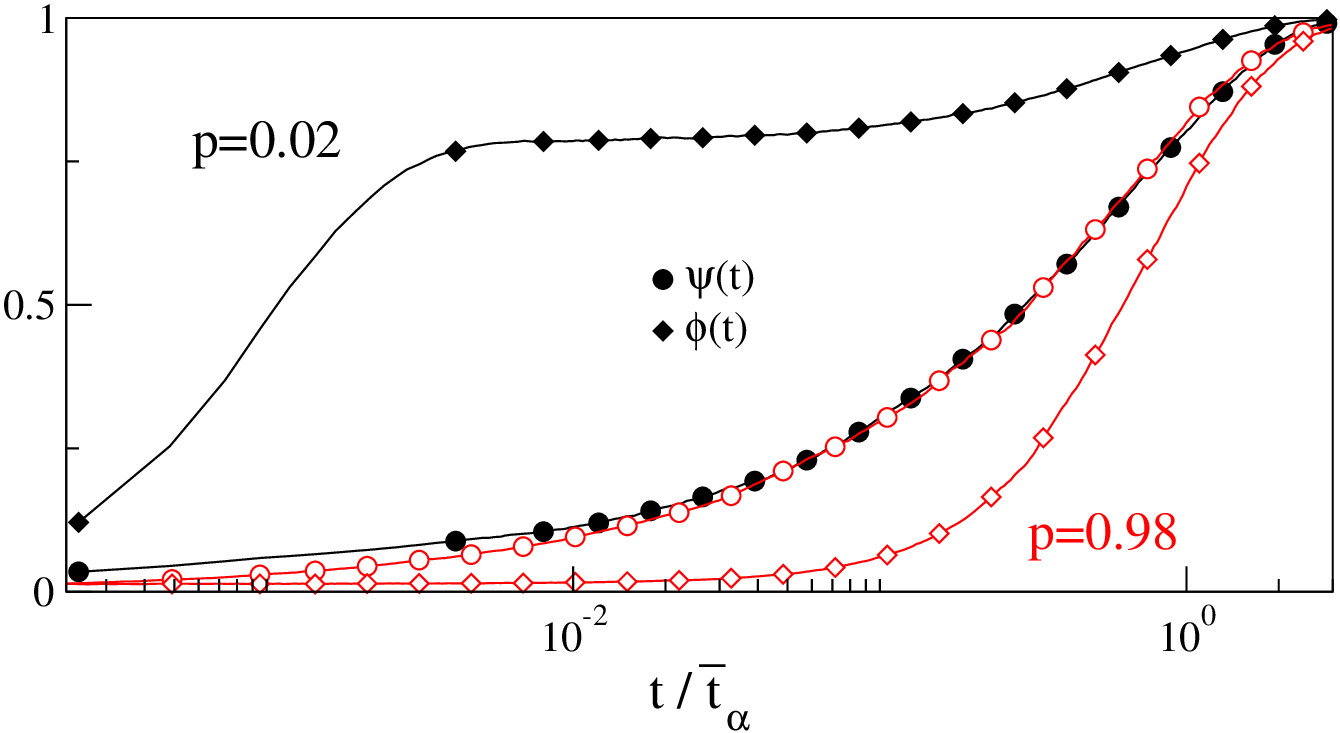}
  \end{center}
   \caption{Ordering mechanisms as a function of
     mobility. In the limit $ p \to 0$ (black curves) local
     order ($\phi(t)$) grows in short time, while global order
     ($\psi(t)$) emerges as a result of vertex-vertex
     competition. When $p \to 1$ (red curves), on the other
     hand, local order emerges only as a result of global ordering at
     late times. Data refer to a fully connected network with $V=100$
     with homogeneous initial conditions $\rho^{+1}(0)=\rho^{-1}(0)=10$.}
  \label{f:op}
\end{figure}

The asymmetry of the fixation time for $p_{+1}=p_{-1}$,
Eq.~(\ref{eq:14}), hints towards different mechanisms in operation on
the way in which convergence is reached in the two limits $p \to 0$ and
$p \to 1$ \cite{takahata91:_geneal}.  To quantify this intuition
it is helpful to consider the global order parameter $\psi(t) =
|N_{+1}(t)-N_{-1}(t)|/N(t),$ measuring the global difference between
the number of individuals belonging to the two species, and the local
order parameter $\phi(t)$ given by the fraction of vertices in which a
local convergence has been reached and only one species is present.
Fig.~\ref{f:op} shows the behavior of these quantities.  When $p$ is
small, intra-vertex order rapidly emerges but different species
prevail in different vertices, as reflected by the low value of the
global order parameter.  The process then proceeds through a
vertex-vertex competition leading in the end to global convergence
thanks to the successive contamination of different vertices.  When
particle mobility is high, on the other hand, convergence emerges
instead by the sudden prevalence of one of the two species, so that
local and global order rise almost simultaneously.

In conclusion, we have studied a metapopulation scheme that allows to
consider the effects of mobility in ordering dynamics.  Focusing on
the Voter/Moran processes as simple yet paradigmatic examples, we have
found expressions for the fixation probability and time, which are
independent from the topological details of the underlying
network. While the fixation probability takes the same form as in the
usual fermionic counterparts, the fixation time depends strongly on
mobility when all species share the same mobility ratio (actually
diverging when the mobility tends to very large or small values).
Additionally, in this regime we have identified two different
mechanisms leading to local and global convergence in the limit of low
and high diffusion ratios. 
Our work opens the way to a better understanding of
mobility in a wide class of models of ordering dynamics, with
consequences touching the broad spectrum of disciplines that have
borrowed from this field over time. In particular, a challenging task
for future work will consider the implementation of mobility in more
complex and realistic models of social dynamics
\cite{castellano2007sps}.

\vspace*{0.5cm}

\textit{Acknowledgments} We acknowledge financial support from the
Spanish MEC (FEDER), under project No. FIS2007-66485-C02-01, and
through ICREA Academia, funded by the Generalitat de Catalunya.  A. B.
acknowledges support of Spanish MCI through the Juan de la Cierva
program.


\providecommand{\newblock}{}

\end{document}